\documentclass[twocolumn,showpacs,preprintnumbers,amsmath,amssymb,prc]{revtex4-1}
\usepackage{epsfig}

\usepackage{mathptmx, courier, pifont}
\usepackage[scaled=0.92]{helvet}
\usepackage[T1]{fontenc}
\usepackage{textcomp}

 \def\ket{\!>\,} \def\ack{\,|\,}
\begin{document}


\author{J. A.~Sheikh$^{1,2}$, G. H.~Bhat$^{1}$, Yan-Xin Liu$^{3,4,5}$,
Fang-Qi Chen$^{6}$, Yang Sun$^{6,3,2} \footnote{Corresponding author
at SJTU: sunyang@sjtu.edu.cn}$ }

\address{$^1$Department of Physics, University of Kashmir, Srinagar,
190 006, India \\
$^2$Department of Physics and Astronomy, University of
Tennessee, Knoxville, TN 37996, USA\\
$^3$Institute of Modern Physics, Chinese Academy of Sciences,
Lanzhou 730000, People's Republic of China\\
$^4$Graduate University of Chinese Academy of Sciences, Beijing
100049, People's Republic of China\\
$^5$School of Science, Huzhou Teachers College, Huzhou 313000,
People's Republic of China \\
$^6$Department of Physics, Shanghai Jiao Tong University, Shanghai
200240, People's Republic of China }

\title{Mixing of quasiparticle excitations and $\gamma$-vibrations
in transitional nuclei}

\date{\today}

\begin{abstract}
Evidence of strong coupling of quasiparticle excitations with
$\gamma$-vibration is shown to occur in transitional nuclei.
High-spin band structures in $^{166,168,170,172}$Er are studied by
employing the recently developed multi-quasiparticle triaxial
projected shell model approach.  It is demonstrated that a low-lying
$K=3$ band observed in these nuclei, the nature of which has
remained unresolved, originates from the angular-momentum projection
of triaxially deformed two-quasiparticle (qp) configurations.
Further, it is predicted that the structure of this band depends
critically on the shell filling: in $^{166}$Er the lowest $K=3$ 2-qp
band is formed from proton configuration, in $^{168}$Er the $K=3$
neutron and proton 2-qp bands are almost degenerate, and for
$^{170}$Er and $^{172}$Er the neutron $K=3$ 2-qp band becomes
favored and can cross the $\gamma$-vibrational band at high
rotational frequencies.  We consider that these are few examples in
even-even nuclei, where the three basic modes of rotational,
vibrational, and quasi-particle excitations co-exist close to the
yrast line.
\end{abstract}


\maketitle

Major advances in experimental techniques have made it feasible to
perform detailed measurements of atomic nuclei at the extremes of
angular-momentum, isospin, and stability.  Detailed spectroscopic
studies have provided deep insights in our understanding of nuclear
many-body problem.  Band structures in some nuclei have been
observed with many bands and up to extremely high angular-momentum.
With the near-completion of the new advanced kind of gamma-ray
detector GRETINA in USA, one would expect a vast amount of
high-quality data covering the regions that have never been reached
before.

The classification and the interpretation of the rich band
structures is a challenge to nuclear theory.  The three basic modes
of excitations of rotational, vibrational, and quasi-particle
constitute the primary origin of the observed bands in nuclei
\cite{BM75}.  In spherical nuclei, the energy spectrum is primarily
built on the quasi-particle excitations, while as in well deformed
nuclei rotational bands are observed and are classified using the
Nilsson scheme.  On the other hand, in transitional nuclei the
excitation spectrum is quite rich and intricate and depict interplay
of all the three modes of excitations
\cite{AB82,BR76,BJ91,GU95,AG96}.

Rotational bands built on vibrations in $\beta$- and $\gamma$-degree
of deformation are observed in many transitional nuclei.  In
particular, well-developed $\gamma$-bands are known to exist in most
of the transitional regions of the nuclear chart and a considerable
effort has been devoted to understand the detailed structure of
these bands.  These bands are traditionally interpreted in the
phonon picture with the observed $K=2$ and 4 bands built on one- and
two-$\gamma$-phonon excitations \cite{XA94,XA93,DG94,PE97,NM99}.
Recently, these bands have been re-interpreted using the microscopic
triaxial projected shell model (TPSM) \cite{JS99,YK00,JY01,YJ02}. It
has been demonstrated that three-dimensional projection of angular
momentum from the triaxially-deformed vacuum state of an even-even
system leads to $K=0,2$ and 4 bands that correspond to the ground-,
$\gamma$-, and $\gamma\gamma$-bands observed in nuclei. In a more
recent development \cite{GH08,JG09,JG10,Gao06}, the TPSM approach
has been generalized to include quasiparticle (qp) excitations, and
it was demonstrated that the projection from triaxially-deformed qp
states can result into various excited bands.  These are the
structures that couple $\gamma$-vibration to qp-excitations, based
on which rotational bands are built.  Thus, these bands have
characteristics of all three excitation modes in nuclei and are,
therefore, the best places to show up the interplay among them.
These recent developments in TPSM approach have greatly enhanced the
model predictability and may provide new insights into the observed
bands with unknown structures.  As a matter of fact, by using this
approach, the interpretation of complicated band structures has
reached a quantitative level \cite{Yeo11,Liu11}.

In $^{168}$Er and  $^{170}$Er, well-developed $K=3$ bands have been
observed that are populated as intensively as $\gamma$-bands
\cite{CD00,BS95,EB91,VS94}.  The $K=3$ band is placed between the
$K=2$ and $K=4$ bands, and, as a matter of fact, crosses the
$\gamma$ band in $^{170}$Er at about $I=12$ and becomes quite low in
energy.  The structure of these bands has remained unresolved and
the purpose of the present work is to shed light on the origin of
these low-lying $K=3$ bands.  It is demonstrated, using the
generalized TPSM approach \cite{GH08,JG09}, that these bands are
examples of qp-excitations that are admixed with $\gamma$-vibration,
and their correct description critically depends on the choice of
the basis deformation.  In the present work, we also evaluate the
intra- and inter-band electromagnetic transition probabilities, and
it is shown that the deformation used in the present work provides a
better agreement for the transition calculations in comparison to
our earlier work on the ground-state configuration only \cite{PJ02}.
The TPSM approach has already been discussed in our earlier
publications \cite{JS99,YK00,JY01,YJ02,GH08,JG09}, and in the
following we shall provide only a few details of the model that are
relevant to the discussion of the results.

For even-even systems, the TPSM basis are composed of projected 0-qp
state (or qp-vacuum $\ack\Phi\ket$), 2-proton, 2-neutron, and 4-qp
configurations, i.e.,
\begin{equation}
\begin{array}{r}
\hat P^I_{MK}\ack\Phi\ket;\\
~~\hat P^I_{MK}~a^\dagger_{p_1} a^\dagger_{p_2} \ack\Phi\ket;\\
~~\hat P^I_{MK}~a^\dagger_{n_1} a^\dagger_{n_2} \ack\Phi\ket;\\
~~\hat P^I_{MK}~a^\dagger_{p_1} a^\dagger_{p_2}
a^\dagger_{n_1} a^\dagger_{n_2} \ack\Phi\ket ,
\label{basis}
\end{array}
\end{equation}
where the three-dimensional angular-momentum operator \cite{KY95}
is given by
\begin{equation}
\hat P^I_{MK} = {2I+1 \over 8\pi^2} \int d\Omega\,
D^{I}_{MK}(\Omega)\, \hat R(\Omega),
\label{PD}
\end{equation}
with $\hat R(\Omega)$ being the rotation operator and
$D^{I}_{MK}(\Omega)$ the $D$-function.  The qp states are obtained
by usual BCS calculations for the deformed single-particle states.
Particle number is conserved on average through the introduction of
the first order Lagrange multipliers.  The values of the
corresponding neutron and proton chemical potentials are obtained by
the constraint for given neutron and proton numbers of nuclei under
consideration.  This ensures the correct shell filling \cite{KY95}.

It is important to note that for the case of axial symmetry, the
qp-vacuum state has $K=0$ \cite{KY95}, where as in the present case
of triaxial deformation, the vacuum state $\ack\Phi\ket$, as well as
any configuration in (\ref{basis}),  is a superposition of all
possible $K$-values. Rotational bands with the triaxial basis states
in (\ref{basis}) are obtained by specifying different values for the
$K$-quantum number in the angular-momentum projector in Eq.
(\ref{PD}).  The allowed values of the $K$-quantum number for a
given intrinsic state are obtained through the following symmetry
consideration. For $\hat S = e^{-\imath \pi \hat J_z}$, we have
\begin{equation}
\hat P^I_{MK}\left|\Phi\right> = \hat P^I_{MK} \hat S^{\dagger} \hat S
\left|\Phi\right> = e^{\imath \pi (K-\kappa)}
\hat P^I_{MK}\left|\Phi\right>.
\label{condition}
\end{equation}
For the self-conjugate vacuum or 0-qp state, $\kappa=0$ and,
therefore, it follows from the above equation that only $K=$ even
values are permitted for this state. For 2-qp states, $a^\dagger
a^\dagger \left|\Phi\right>$, the possible values for $K$-quantum
number are both even and odd, depending on the structure of the
qp-state. For example, for a 2-qp state formed from the combination
of the normal and the time-reversed states having $\kappa = 0$, only
$K$ = even values are permitted. For the combination of the two
normal states, $\kappa=1$, only $K=$ odd states are permitted.

As in the earlier PSM calculations, we use the pairing plus
quadrupole-quadrupole Hamiltonian \cite{KY95}
\begin{equation}
\hat H = \hat H_0 - {1 \over 2} \chi \sum_\mu \hat Q^\dagger_\mu
\hat Q^{}_\mu - G_M \hat P^\dagger \hat P - G_Q \sum_\mu \hat
P^\dagger_\mu\hat P^{}_\mu,
\label{hamham}
\end{equation}
with the last term in (\ref{hamham}) being the quadrupole-pairing
force.  The corresponding triaxial Nilsson mean-field Hamiltonian,
which  is obtained by using the Hartree-Fock-Bogoliubov (HFB)
approximation, is given by
\begin{equation}
\hat H_N = \hat H_0 - {2 \over 3}\hbar\omega\left\{\epsilon\hat Q_0
+\epsilon'{{\hat Q_{+2}+\hat Q_{-2}}\over\sqrt{2}}\right\}.
\label{nilsson}
\end{equation}
Here $\hat H_0$ is the spherical single-particle Hamiltonian, which
contains a proper spin-orbit force \cite{Ni69}. The interaction
strengths are taken as follows:  The $QQ$-force strength $\chi$ is
adjusted such that the physical quadrupole deformation $\epsilon$ is
obtained as a result of the self-consistent mean-field HFB
calculation \cite{KY95}.  The monopole pairing strength $G_M$ is of
the standard form
\begin{equation}
G_{M} = \left(G_{1}\mp G_{2}\frac{N-Z}{A}\right)\frac{1}{A}
\,(\rm{MeV}), \label{gmpairing}
\end{equation}
where $- (+)$ is for neutron (proton). In the present calculation,
we use $G_1=20.12$ and $G_2=13.13$, which approximately reproduce
the observed odd-even mass difference in this region.  This choice
of $G_M$ is appropriate for the single-particle space employed in
the model, where three major shells are used for each type of
nucleons ($N=3,4,5$ for protons and $N=4,5,6$ for neutrons).  The
quadrupole pairing strength $G_Q$ is assumed to be proportional to
$G_M$, and the proportionality constant being fixed as 0.16.  These
interaction strengths are consistent with those used earlier for the
same mass region \cite{JS99,YK00,KY95}.  Deformation parameters used
to construct the qp-basis are listed in Table I.

\begin{table}
\caption{The deformation parameters used in the calculation for
$^{166,168,170,172}$Er.  The axial deformation $\epsilon$ is taken
from Ref. \cite{Raman} (converted from $\beta$ values given there to
$\epsilon$ by multiplying 0.95 factor). The triaxial deformation
parameter is denoted by $\epsilon'$.}
\begin{tabular}{|ccccc|}
\hline             & $^{166}$Er & $^{168}$Er & $^{170}$Er & $^{172}$Er \\
\hline  $\epsilon$ & 0.325      & 0.321      & 0.319      & 0.314      \\
        $\epsilon'$& 0.126      & 0.125      & 0.110      & 0.110      \\\hline
\end{tabular}
\end{table}

\begin{figure}[htb]
 \centerline{\includegraphics[trim=0cm 0cm 0cm
0cm,width=0.40\textwidth,clip]{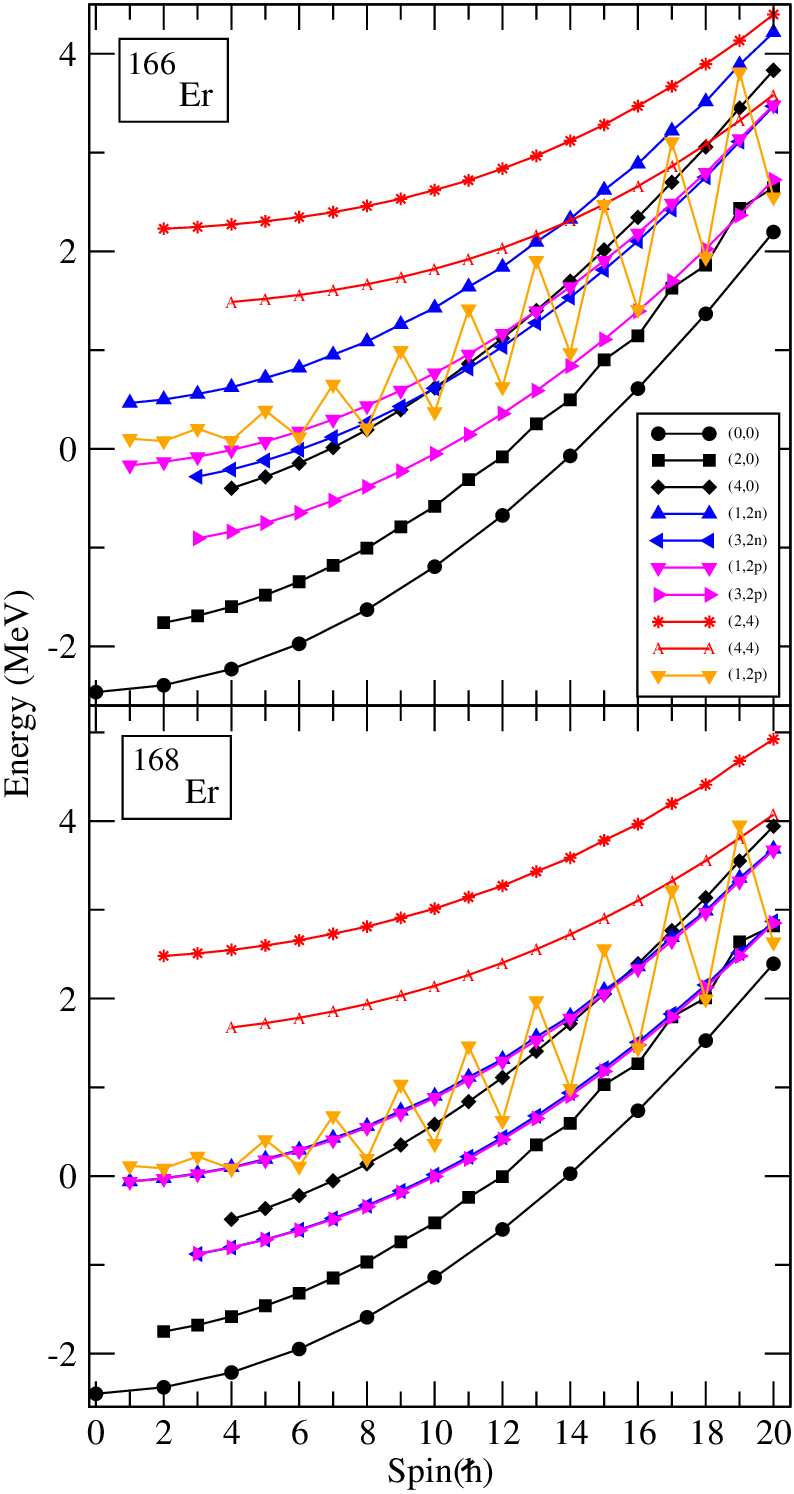}} \caption{(Color
online) Band diagrams for $^{166-168}$Er isotopes. The labels
$(K$,n-qp) indicate the $K$-value and the quasiparticle character of
the configuration, for instance, $(3,2p)$ corresponds to the
two-proton configuration with $K=3$. } \label{fig0}
\end{figure}
\begin{figure}[htb]
 \centerline{\includegraphics[trim=0cm 0cm 0cm
0cm,width=0.40\textwidth,clip]{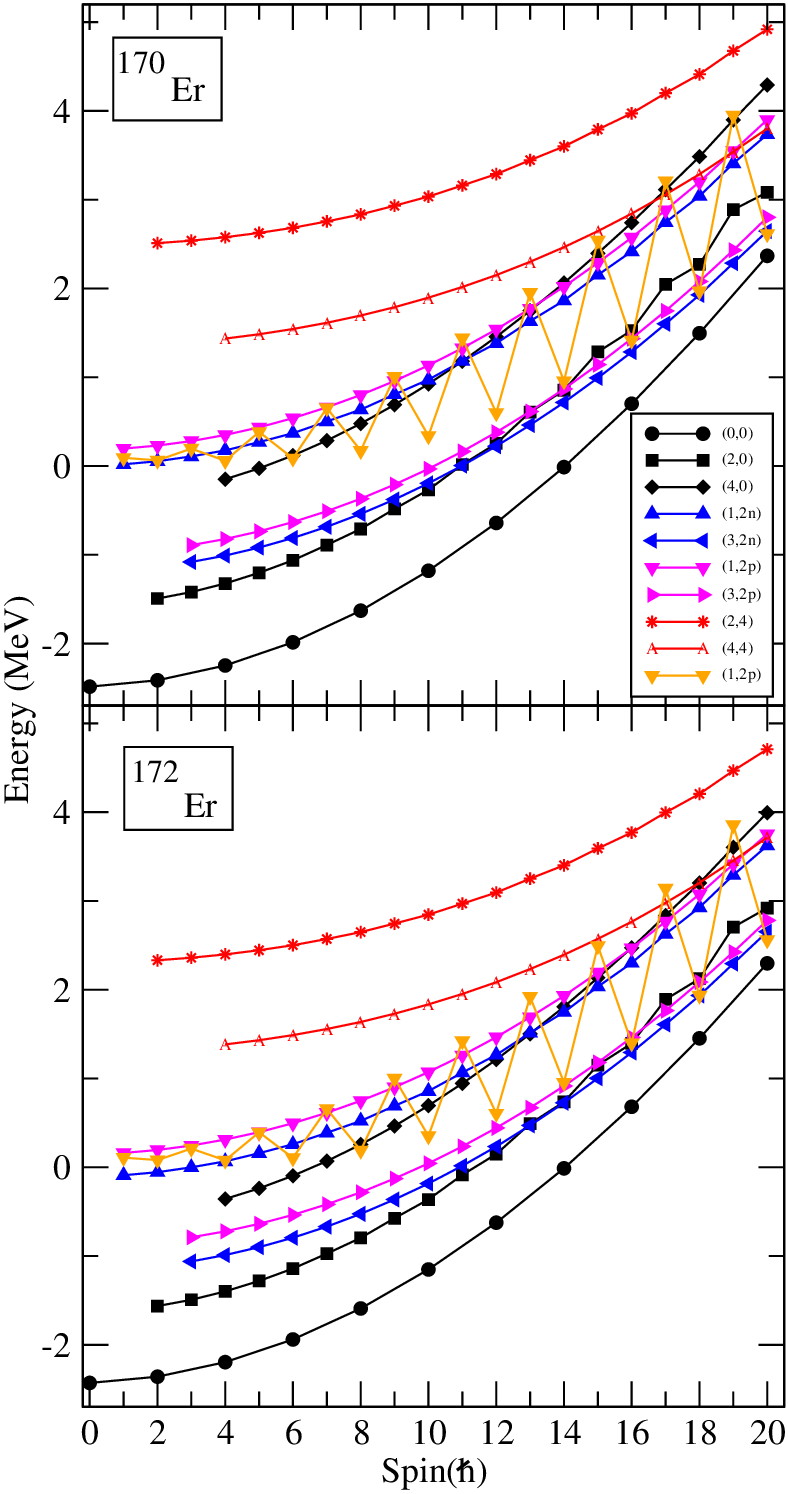}} \caption{(Color
online) Same as in Fig. (\ref{fig0}), but for $^{170-172}$Er
isotopes. } \label{fig1}
\end{figure}

The angular-momentum projected energies from 0-qp, 2-qp, and 4-qp
configurations, calculated with deformation parameters given above,
are depicted in Figs.~\ref{fig0} and \ref{fig1} for the four
Er-isotopes studied in the present work. These are the so-called
band diagrams, defined in the projected shell model \cite{KY95}
approach.  In these figures, the projected energies of only the
lowest few 2- and 4-qp configurations are plotted for clarity.  It
has been already stated that the admissible $K$-values for the
triaxial vacuum state are $K=0,2,4,...$ and the projection from
these possible values give rise to the ground-state band with $K=0$,
$\gamma$-band with $K=2$, $\gamma\gamma$-band with $K=4$, and {\it
etc}.  The calculated unperturbed band heads of $\gamma$- and
$\gamma\gamma$-bands are roughly at energies of (relative to the
ground-state) 0.7033, 2.0643 for $^{166}$Er, 0.7009, 1.9651 for
$^{168}$Er, 0.8659, 2.0730 for $^{170}$Er, and 0.8659, 2.073 for
$^{172}$Er (all in MeV).  The anharmonicity in $\gamma$-vibration
appears automatically from the calculations and correctly describes
the experimental data \cite{YK00,CF96,TH98}.

The projected bands from 2-qp states result into both even- and
odd-$K$ values depending on the combination of the qp-states.  The
bands with $K=1,3,...$ are obtained by combining two normal states
and are traditionally referred to as aligned bands. In many nuclei
in the rare-earth region these aligned bands cross the ground-state
band, giving rise to the phenomenon of backbending.  Although, for
the four Erbium isotopes studied in the present work, these aligned
bands do not cross the ground-state band, they are noted to follow
$\gamma$-bands very closely and interact with them.  It is quite
interesting to observe from Figs.~\ref{fig0} and \ref{fig1} that for
$^{166}$Er, proton 2-qp band with $K=3$ is lower than the
corresponding neutron band, for $^{168}$Er the two bands are nearly
degenerate, and in cases of $^{170,172}$Er the neutron band is lower
than the proton band.  The relative change in the $K=3$ band
structures is attributed to the shell filling of neutrons and
protons. As neutron number increases, the neutron Fermi level
changes, while the proton Fermi level remains almost unchanged for
the isotopes.  The proton and neutron character of the bands can be
probed through measurement of g-factors of the bands.  Four-qp bands
in Figs.~\ref{fig0} and \ref{fig1} are observed to lie higher, and
do not become yrast up to the highest angular-momentum state studied
in the present work.

\begin{figure}[htb]
 \centerline{\includegraphics[trim=0cm 0cm 0cm
0cm,width=0.40\textwidth,clip]{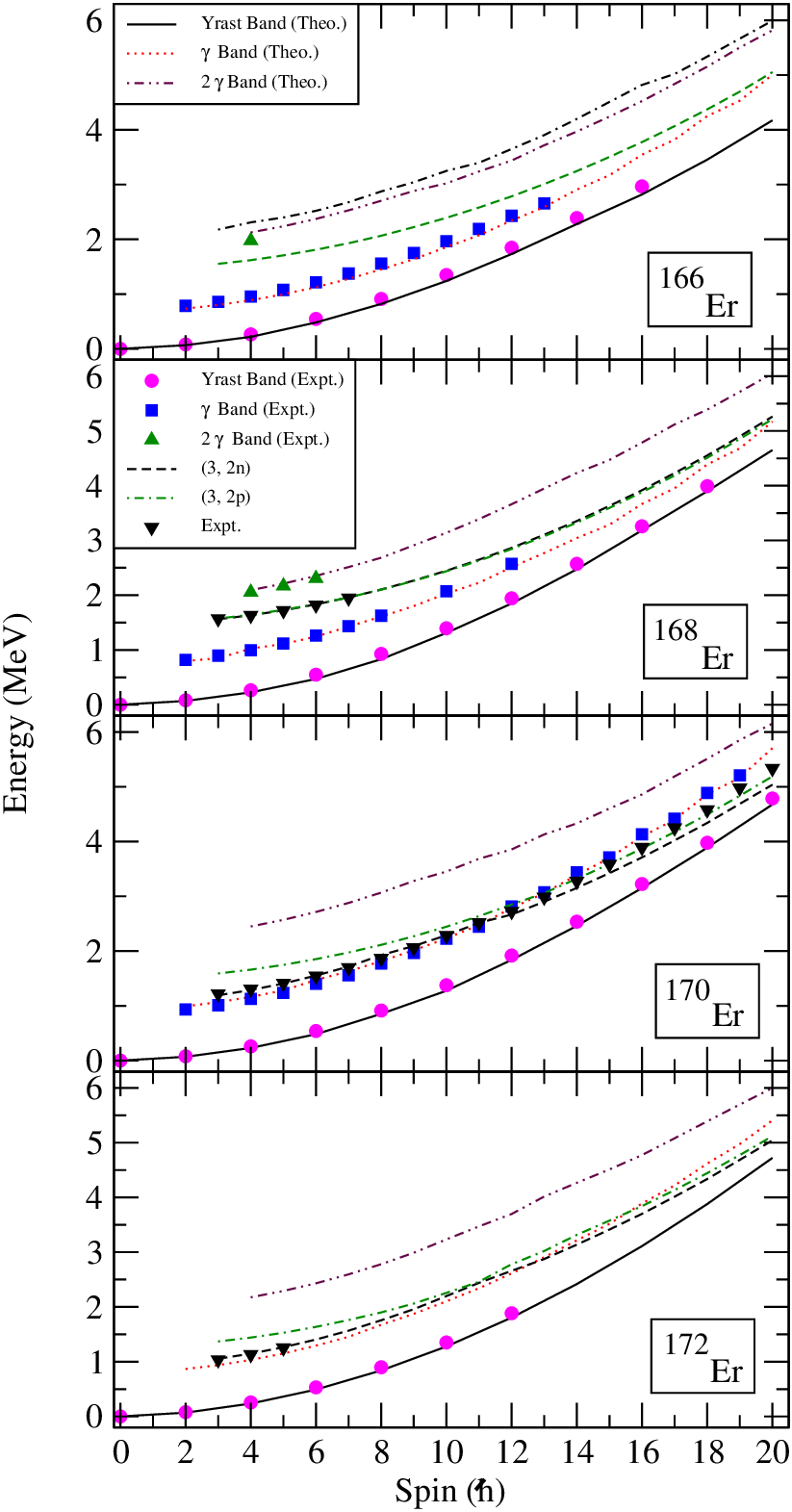}} \caption{(Color
online) Comparison of the TPSM energies after configuration mixing
with the available experimental data for $^{168-172}$Er. Data are
taken from \cite{CD00,CF96,TH98,GD10}.} \label{fig2}
\end{figure}

In the second stage, the projected bands, obtained above, are then
mixed through diagonalization of the shell model Hamiltonian in
(\ref{hamham}).  In band diagrams of Figs.~\ref{fig0} and \ref{fig1}
only the lowest bands were shown, but in the diagonalization process
the projected states employed is nearly 40 for all nuclei.
Fig.~\ref{fig2} depicts the calculated bands after diagonalization
and also displays the corresponding available experimental data.  It
is important to point out that, although the bands in
Fig.~\ref{fig2} are labeled as $\gamma$-, $\gamma\gamma$-, and
$K=3$-bands, these are only the dominant components in the
wavefunction.  The projected states after diagonalization are in
general mixed.  In particular, 2-qp $K=3$ band has a significant
contribution from 0-qp $K=2$ configuration at higher
angular-momenta.

For $^{166}$Er, the agreement between the TPSM and the experimental
energies for the yrast- and the $\gamma$-bands is exceedingly good.
There is only $I=4$ bandhead state known for the $\gamma\gamma$-band
\cite{CF96} and it is also reproduced quite well.  It is noted from
Fig.~\ref{fig2} that the $K=3$ band, which is a projected band from
2-qp proton configuration, is predicted above the known
$\gamma$-band but is lower than the $\gamma\gamma$-band.  We hope
that future high-spin experimental studies shall be able to populate
this band. In the lower panels of Fig.~\ref{fig2}, the results for
the other three studied isotopes also display a good agreement with
the available experimental data.  In $^{168}$Er, the known
experimental data for the $\gamma\gamma$-band \cite{TH98} are also
described correctly.  The 2-qp proton and neutron $K=3$ bands are
almost degenerate for $^{168}$Er, and the observed five states of
this band are noted to be reproduced quite well.  The interesting
prediction is that there are two almost identical $K=3$ bands that
have predominantly proton and neutron structures, respectively.

\begin{figure}[htb]
 \centerline{\includegraphics[trim=0cm 0cm 0cm
0cm,width=0.40\textwidth,clip]{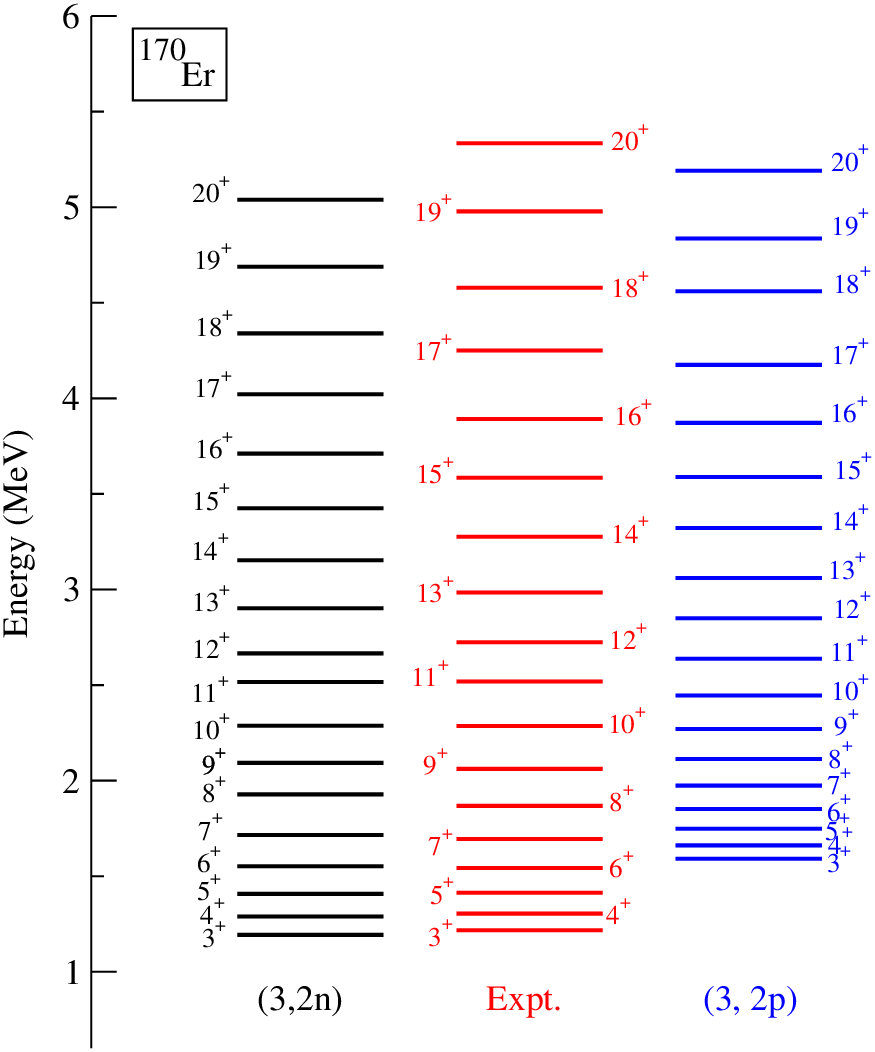}} \caption{(Color
online) Detailed comparison of the calculated $K=3$ bands in
$^{170}$Er with experimental data \cite{CD00}.} \label{fig3}
\end{figure}

In $^{170}$Er, the $K=3$ band is populated as intensively as the
$\gamma$-band and is known up to $I=20$ \cite{CD00}.  Furthermore,
this band crosses the $\gamma$-band at $I=12$ and becomes the first
excited band above this spin value.  The present work reproduces
these properties and what is more interesting is that the observed
small staggering in the $\gamma$-band at higher spins is also borne
out by the TPSM results.  The $K=3$ proton 2-qp band also crosses
the $\gamma$-band and becomes the second excited band above $I=14$.

Fig.~\ref{fig3} presents a more detailed comparison of the observed
and the calculated $K=3$ bands for $^{170}$Er.  The calculated $K=3$
2-qp neutron band agrees quite well with the experimental band,
however, at the top of the band some descrepancies are quite
evident.  There could be several reasons for these discrepancies.
The bulk of the discrepancy could be attributed to the fixed
mean-field assumed in the present study.  The Nilsson potential is
chosen for the mean-field and is determined by the input deformation
parameters, $\epsilon$ and $\epsilon'$.  The pairing potential, on
the other hand, is obtained from the monopole interaction using the
BCS ansatz.  In a more accurate self-consistent treatment,
projection before variation, the mean-field and the pairing
potential are known to vary with qp-excitation and angular-momentum.
Very similar to $^{170}$Er, the heavier isotope $^{172}$Er is also
predicted to exhibit a band crossing picture between the $K=3$ band
\cite{GD10} and the $\gamma$-band.  The two bands follow very
closely for the entire spin region, and interact with each other.
Thus the $K=3$ band in both transitional $^{170}$Er and $^{172}$Er
nuclei, although they have two-quasiparticle structure, interact
strongly with the $\gamma$-vibration.

\begin{figure}[htb]
 \centerline{\includegraphics[trim=0cm 0cm 0cm
0cm,width=0.40\textwidth,clip]{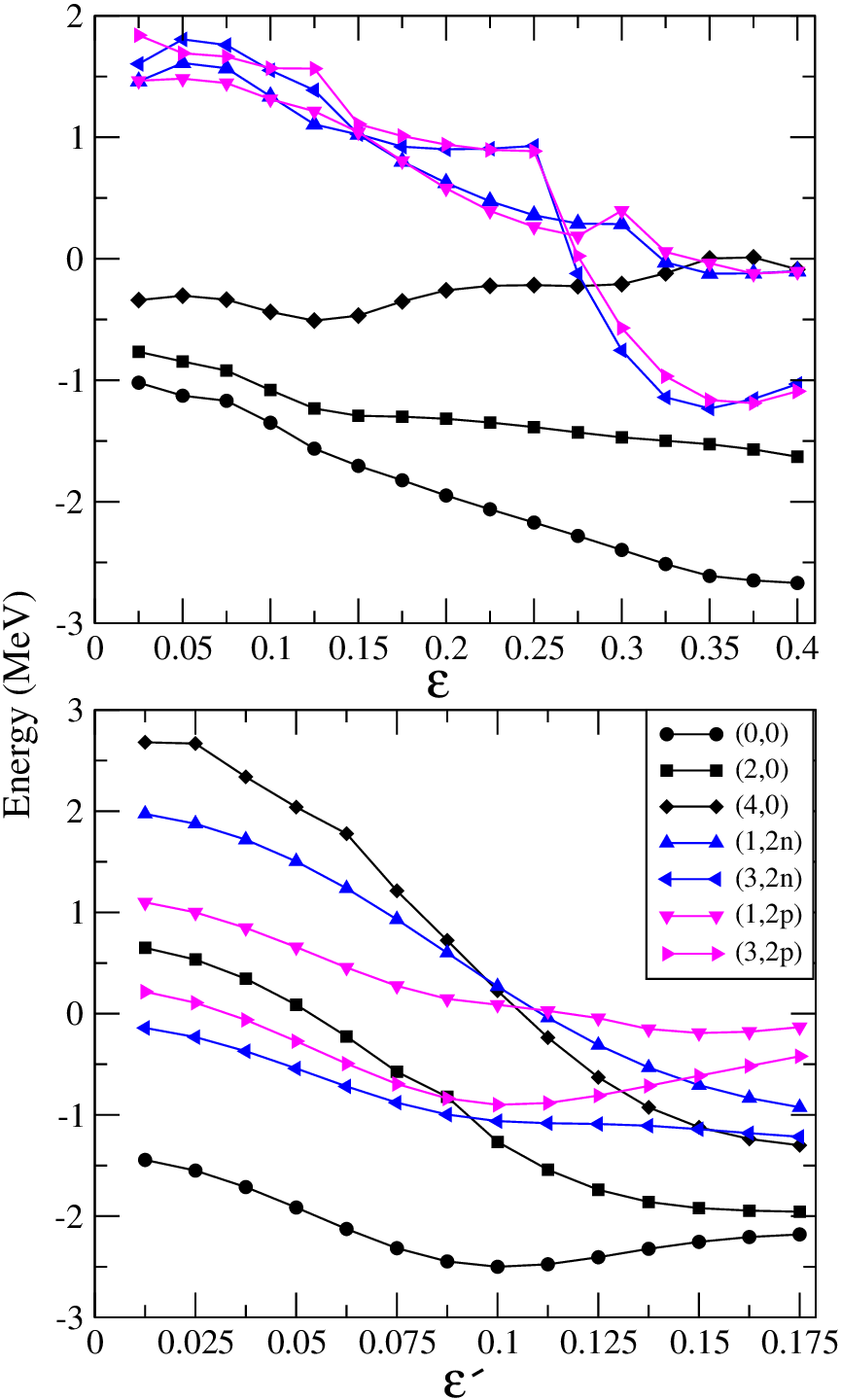}} \caption{(Color
online)  Behavior of the projected energies of various
configurations as a function of axial and triaxial deformations for
$^{170}$Er. In the upper panel, the projected energies have been
evaluated for a fixed value of $\epsilon'=0.11$ and in the lower
panel $\epsilon=0.319$ has been chosen.} \label{fig4}
\end{figure}
\begin{figure}[htb]
 \centerline{\includegraphics[trim=0cm 0cm 0cm
0cm,width=0.35\textwidth,clip]{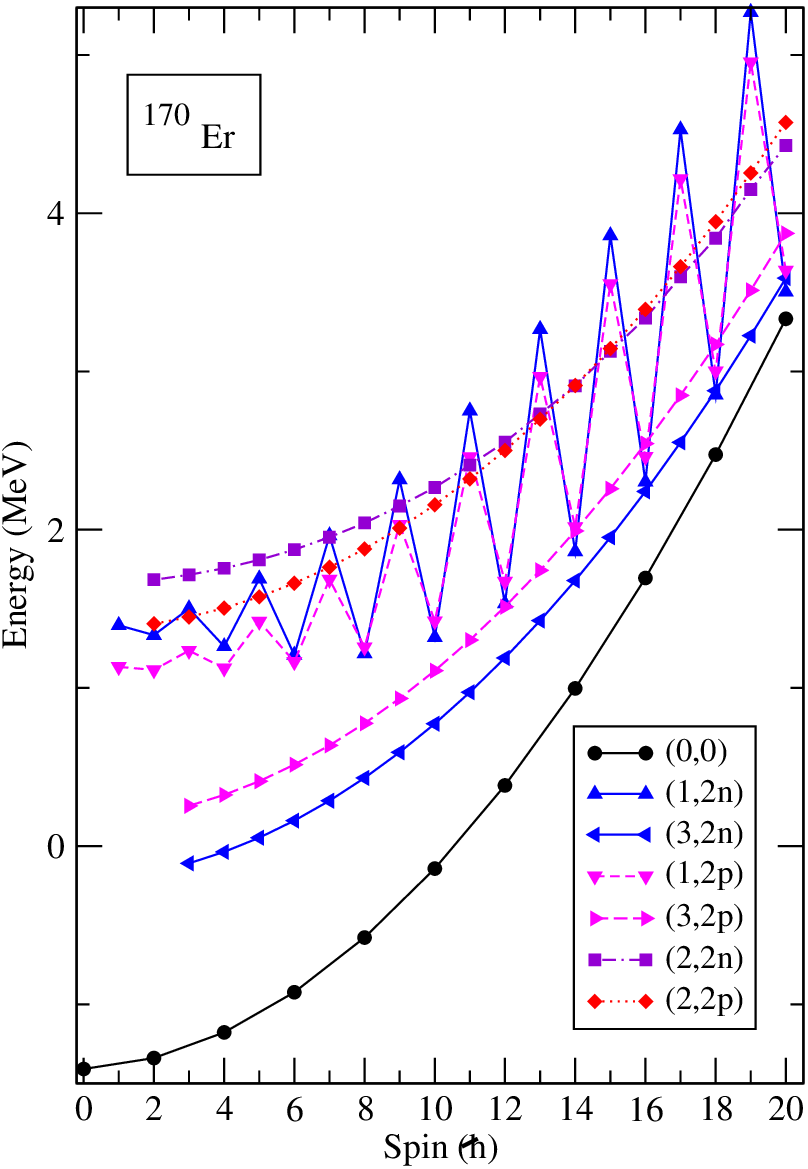}}
\caption{(Color online) Band diagrams for $^{170}$Er by using Axial
Projected Shell Model code \cite{SH97}.} \label{fig5}
\end{figure}

To probe the mechanism behind the appearance of the $K=3$ band,
close to the yrast line, in the Er-isotopes, we have studied the
behavior of the projected energies as a function of deformation
parameters, $\epsilon$ and $\epsilon'$.  As an example, the
variation of the projected energies are shown in Fig.~\ref{fig4} for
$^{170}$Er.  In the upper panel of the figure, the variation is
depicted as a function of the axial deformation $\epsilon$ with
fixed triaxial deformation $\epsilon'=0.11$.  For low axial
deformation values, the $K=3$ band is higher than the $K=1$ band.
However, above $\epsilon=0.25$ $K=3$ band depicts a large downward
shift and becomes lower than the $K=1$ and $\gamma\gamma$-band.  The
other bands are noted to be less sensitive to the axial deformation.
The dependence of the projected energies on $\epsilon'$, shown in
the lower panel of Fig.~\ref{fig4}, is calculated for fixed
$\epsilon=0.319$.  This dependence, first of all, clearly
demonstrates that the ground-state minimum has $\epsilon'=0.11$ and
this value has been considered in all our earlier calculations. This
figure also shows that the $K=3$ band is less sensitive to
$\epsilon'$ as compared to the other bands.  Therefore, it can be
concluded from the present results that the appearance of the
low-excitation $K=3$ band is primarily due to the axial deformation.

The question obviously arises on the relevance of the triaxial
deformation in the structure of $K=3$ bands.  To investigate this
question, we have also performed axial projected shell model
calculations using the original projected shell model code
\cite{SH97}.  The parameters used in this study are exactly same as
those used in the above triaxial study, except that now the triaxial
deformation is absent. The advantage of the axial study is that it
provides a direct information on the $K$-structure of the
quasi-particle states.  The results of the axial study are plotted
in Fig.~\ref{fig5}.  The ground-state band with $K=0$ is the
projection from the qp-vacuum state with axial symmetry, and all
other bands in Fig.~\ref{fig5} are the projected bands from the 2-qp
states.  The lowest 2-qp band is the neutron band with $K=3$ and is
formed from the Nilsson states of $[651]1/2$ and $[633]7/2$,  and
proton 2-qp band is formed from $[523]7/2$ and $[541]1/2$. Although,
the $K=3$ band is lower in this axial case as well, the obtained
band structures are completely different from those in the triaxial
study. First of all, as expected, there is no $\gamma$-band as in
the triaxial case and also in the observed data.  Secondly, the band
head of the $K=3$ band  is higher in the axial case as compared to
the corresponding experimental band head.  Therefore, although the
axial deformation is important for the $K=3$ band to appear, the
strong mixing with $\gamma$-degree of freedom is crucial to explain
its excitation and the rotational behavior.

In the present work, we have also evaluated the B(E2) transition
probabilities, which are presented in Tables II along the yrast
bands for the studied isotopes.  Further, we calculated the
inter-band transitions between $K=3$ and $\gamma$-bands for
$^{170,172}$Er as these two bands cross for these isotopes, and the
transitions are displayed in Table III.  The B(E2) values have been
calculated using the standard effective charges of $e_p=1.5e$ and
$e_n=0.5e$.  It is evident from Table II that calculated B(E2)
describe the known transitions well.  For the inter-band transitions
in $^{170,172}$Er between the $K=3$ and $K=2$ band shown in Table
III, it is interesting to note that in the crossing region of the
two bands (around spin 12), very enhanced B(E2) is predicted.  The
large inter-band B(E2) values indicate a considerable overlap
between the wavefunctions, implying a strong mixing between the
quasiparticles and the $\gamma$-vibration.

In summary, we have presented evidences for strong coupling of
quasiparticle excitations with $\gamma$-vibration in transitional
nuclei.  High-spin band structures in a series of Erbium isotopes
$^{166,168,170,172}$Er have been studied by using the recently
developed multi-quasiparticle triaxial projected shell model
approach.  The spotlight of the present investigation has been the
$K=3$ band observed in some of these nuclei that is populated as
strongly as the $\gamma$-band.  In the framework of triaxial
angular-momentum projection, we have shown that this band has mainly
a structure of triaxially deformed 2-qp state projected to the $K=3$
component. This is to compare with the traditional $\gamma$- and
$\gamma\gamma$-bands, which are based on triaxially-deformed 0-qp
state projected to the $K=2$ and 4 components, respectively.  It has
been further shown that the detailed structure and position of the
$K=3$ band depend sensitively on the shell filling.  In $^{166}$Er
the lowest $K=3$ 2-qp band is formed from proton configuration, in
$^{168}$Er the $K=3$ neutron and proton 2-qp bands are almost
degenerate, and for $^{170,172}$Er the neutron $K=3$ 2-qp band
becomes favored.  The prediction of systematic appearance of two
$K=3$ bands with proton and neutron structures, close to the yrast
line, awaits experimental confirmation.

The calculations presented in the present article have clearly
demonstrated that a simple model based on schematic pairing plus
quadrupole-quadrupole interaction with three-dimensional
angular-momentum projection technique can describe the near yrast
band structures in transitional nuclei in a quantitative manner.  A
drawback in the present analysis is the uncertainty in the strength
parameters of the schematic interaction.  In future studies, we are
planning to adopt a recently developed mapping procedure
\cite{Alh08,Alh06} to microscopically determine the strength
parameters.  In this new approach, the energy surfaces obtained from
the schematic effective interaction with free strength parameters
are optimized to reproduce the energy surfaces retrieved from a
realistic density functional approach.

\begin{widetext}
\begin{table*}
\caption{Comparison of known experimental yrast-band $B(E2)$ values
(in w.u., and associated errors in parenthesis) and calculated ones
for $^{166,168,170,172}$Er isotopes.}
\begin{tabular}{cccccccc}
\hline $(I,K)_i\rightarrow (I,K)_f$ & $^{166}$Er (expt.) &
$^{166}$Er (theo.) &$^{168}$Er (expt.) &
$^{168}$Er (theo.) &$^{170}$Er (expt.) &$^{170}$Er (theo.) &$^{172}$Er (theo.)\\
\hline
$(2,0)_i\rightarrow (0,0)_f$ & 214 (\textit{10)}  & 245.02  &  207 (\textit{6)}  & 242.51   & 208 (\textit{4)}  &  244.96   & 241.07  \\
$(4,0)_i\rightarrow (2,0)_f$ & 311 (\textit{20)}  & 351.63  &  318 (\textit{12)} & 347.77   &                   &  350.83   & 345.59  \\
$(6,0)_i\rightarrow (4,0)_f$ & 347 (\textit{45)}  & 390.38  &  440 (\textit{30)} & 385.67   &                   &  388.15   & 383.01  \\
$(8,0)_i\rightarrow (6,0)_f$ & 365 (\textit{50)}  & 413.07  &  350 (\textit{20)} & 407.62   & 370 (\textit{30)} &  408.88   & 404.47  \\
$(10,0)_i\rightarrow (8,0)_f$ & 371 (\textit{46)} & 429.60  &  302 (\textit{21)} & 423.73   & 320 (\textit{22)} &  423.20   & 419.99  \\
$(12,0)_i\rightarrow (10,0)_f$ &                  & 442.77  &  334 (\textit{22)} & 437.17   & 375 (\textit{20)} &  434.33   & 432.71  \\
$(14,0)_i\rightarrow (12,0)_f$ &                  & 453.01  &                    & 449.09   &                   &  443.35   & 443.66  \\
$(16,0)_i\rightarrow (14,0)_f$ &                  & 459.62  &                    & 459.93   &                   &  450.50   & 453.02  \\
$(18,0)_i\rightarrow (16,0)_f$ &                  & 461.29  &                    & 469.96   &                   &  445.86   & 460.55  \\
$(20,0)_i\rightarrow (18,0)_f$ &                  & 456.51  &                    & 479.35   &                   &  459.95   & 466.01  \\
\hline
\end{tabular}
\end{table*}
\end{widetext}

\begin{table}[t]
\caption{Calculated inter-band $B(E2)$ values (in w.u.) from the
$K=3$ to $\gamma$ band for $^{170,172}$Er isotopes.}
\begin{tabular}{ccc} \hline
$(I,K)_i\rightarrow (I,K)_f$ & $^{170}$Er (theo.) & $^{172}$Er (theo.)\\ \hline
$(4,3)_i\rightarrow (2,2)_f$   & 0.18&   0.01   \\
$(6,3)_i\rightarrow (4,2)_f$   & 0.68 &  0.03    \\
$(8,3)_i\rightarrow (6,2)_f$   & 4.53 &   0.07      \\
$(10,3)_i\rightarrow (8,2)_f$  & 91.27&  0.64   \\
$(12,3)_i\rightarrow (10,2)_f$ & 105.16& 23.33  \\
$(14,3)_i\rightarrow (12,2)_f$ & 40.04 & 110.24   \\
$(16,3)_i\rightarrow (14,2)_f$ & 0.64 &  84.16  \\
$(18,3)_i\rightarrow (16,2)_f$ & 0.29 &  0.23    \\
$(20,3)_i\rightarrow (18,2)_f$ & 0.07 &  0.01  \\
\hline
\end{tabular}
\end{table}

This work was supported in part by the National Natural Science
Foundation of China under contract Nos. 10875077, 11075103, and
10975051, the National Natural Science Foundation of Huzhou under
contract No. 2010YZ11, the Shanghai Pu-Jiang grant, and the Chinese
Academy of Sciences.

\end{document}